\documentclass[fleqn,usenatbib]{mnras}

\usepackage{newtxtext,newtxmath}
\usepackage{pdflscape}
\usepackage[T1]{fontenc}

\DeclareRobustCommand{\VAN}[3]{#2}
\let\VANthebibliography\thebibliography
\def\thebibliography{\DeclareRobustCommand{\VAN}[3]{##3}\VANthebibliography}

\usepackage{scalerel}
\usepackage{tikz}
\usetikzlibrary{svg.path}

\definecolor{orcidlogocol}{HTML}{A6CE39}
\tikzset{
  orcidlogo/.pic={
    \fill[orcidlogocol] svg{M256,128c0,70.7-57.3,128-128,128C57.3,256,0,198.7,0,128C0,57.3,57.3,0,128,0C198.7,0,256,57.3,256,128z};
    \fill[white] svg{M86.3,186.2H70.9V79.1h15.4v48.4V186.2z}
                 svg{M108.9,79.1h41.6c39.6,0,57,28.3,57,53.6c0,27.5-21.5,53.6-56.8,53.6h-41.8V79.1z M124.3,172.4h24.5c34.9,0,42.9-26.5,42.9-39.7c0-21.5-13.7-39.7-43.7-39.7h-23.7V172.4z}
                 svg{M88.7,56.8c0,5.5-4.5,10.1-10.1,10.1c-5.6,0-10.1-4.6-10.1-10.1c0-5.6,4.5-10.1,10.1-10.1C84.2,46.7,88.7,51.3,88.7,56.8z};
  }
}

\newcommand\orcidicon[1]{\href{https://orcid.org/#1}{\mbox{\scalerel*{
\begin{tikzpicture}[yscale=-1,transform shape]
\pic{orcidlogo};
\end{tikzpicture}
}{|}}}}
\usepackage{hyperref}
\usepackage{graphicx}	% Including figure files
\usepackage{amsmath}	% Advanced maths commands
\usepackage{multirow}
\usepackage{float}

\newcommand\msun{M_{\odot}}
\newcommand\mhi{M_{\ion{H}{i}}}
\newcommand\LV{L_{V}}
\newcommand\lsun{L_{\odot}}

\newcommand{\kms}{{\ensuremath{\mathrm{km\,s^{-1}}}}}

\defcitealias{Carlsten2020}{C20}
\defcitealias{Carlsten2021}{C21}
\defcitealias{2021Mao}{M21}
\defcitealias{ELVESI}{C22}
\title[Quenched Satellites Around MW Analogs]{The Quenched Satellite Population Around Milky Way Analogs}

\author[Karunakaran et al.]{Ananthan Karunakaran$^{1\,\orcidicon{0000-0001-8855-3635}}$\,\thanks{E-mail: ananthan@iaa.es}, David J. Sand$^{2\,\orcidicon{0000-0003-4102-380X}}$, Michael G. Jones$^{2\,\orcidicon{0000-0002-5434-4904}}$, Kristine Spekkens$^{3,4\,\orcidicon{0000-0002-0956-7949}}$,
\newauthor{Paul Bennet$^{5\,\orcidicon{0000-0001-8354-7279}}$, Denija Crnojevi\'{c}$^{6\,\orcidicon{0000-0002-1763-4128}}$, Bur\c{c}{\rlap{\.}\i}n Mutlu-Pakd{\rlap{\.}\i}l$^{7\,\orcidicon{0000-0001-9649-4815}}$, Dennis Zaritsky$^{2\,\orcidicon{0000-0002-5177-727X}}$}
\\
$^{1}$Instituto de Astrof\'{i}sica de Andaluc\'{i}a (CSIC), Glorieta de la Astronom\'{i}a, 18008 Granada, Spain\\
$^{2}$Steward Observatory, University of Arizona, 933 North Cherry Avenue, Rm. N204, Tucson, AZ 85721-0065, USA\\
$^{3}${Department of Physics and Space Science, Royal Military College of Canada P.O. Box 17000, Station Forces Kingston, ON K7K 7B4, Canada}\\
$^{4}${Department of Physics, Engineering Physics and Astronomy, Queen's University, Kingston, ON K7L 3N6, Canada}\\
$^{5}${Space Telescope Science Institute, 3700 San Martin Drive, Baltimore, MD 21218, USA}\\
$^{6}${University of Tampa, 401 West Kennedy Boulevard, Tampa, FL 33606, USA}\\
$^{7}${Department of Physics and Astronomy, Dartmouth College, Hanover, NH 03755, USA}\\
}

\date{Accepted XXX. Received YYY; in original form ZZZ}

\pubyear{2022}

\begin{document}
\label{firstpage}
\pagerange{\pageref{firstpage}--\pageref{lastpage}}
\maketitle

\begin{abstract}
We study the relative fractions of quenched and star-forming satellite galaxies in the Satellites Around Galactic Analogs (SAGA) survey and Exploration of Local VolumE Satellites (ELVES) program, two nearby and complementary samples of Milky Way-like galaxies that take different approaches to identify faint satellite galaxy populations.\ We cross-check and validate sample cuts and selection criteria, as well as explore the effects of different star-formation definitions when determining the quenched satellite fraction of Milky Way analogs.\ We find the mean ELVES quenched fraction ($\langle QF\rangle$), derived using a specific star formation rate (sSFR) threshold, decreases from $\sim$50\% to $\sim$27\% after applying a cut in absolute magnitude to match that of the SAGA survey ($\langle QF\rangle_{SAGA}\sim$9\%).\ We show these results are consistent for alternative star-formation definitions.\ Furthermore, these quenched fractions remain virtually unchanged after applying an additional cut in surface brightness.\ Using a consistently-derived sSFR and absolute magnitude limit for both samples, we show that the quenched fraction and the cumulative number of satellites in the ELVES and SAGA samples broadly agree.\ We briefly explore radial trends in the ELVES and SAGA samples, finding general agreement in the number of star-forming satellites per host as a function of radius.\ Despite the broad agreement between the ELVES and SAGA samples, some tension remains with these quenched fractions in comparison to the Local Group and simulations of Milky Way analogs.\
\end{abstract}

\begin{keywords}
galaxies: dwarf -- galaxies: star formation -- galaxies: evolution -- galaxies: formation -- (galaxies:) Local Group
\end{keywords}

\section{Introduction}
Dwarf galaxies play a crucial role in verifying and refining our model of galaxy formation and evolution in the $\Lambda$CDM framework \citep[e.g.][]{2017bullockBoylanKolchin}.\ The Local Group (LG) has long been the foremost laboratory for studies of dwarf galaxies, including the satellites of the two massive hosts, the Milky Way (MW) and M31, as well as those in the periphery beyond the immediate influence of these massive systems.\ While the LG affords the opportunity to study some of the faintest systems, it is only one system.\ In order to have a statistical understanding of the interplay between MW-like hosts and their satellites, we must expand studies beyond the LG. Work in this respect is underway.\

One focus is on how the environment of dwarfs impacts their neutral hydrogen (HI) gas reservoirs and their star formation.\ The vast majority of LG satellites within the virial radius of their host are gas-poor and those beyond it are typically gas-rich \citep{Spekkens2014,Putman2021}.\ Similarly, the quenched fraction of LG satellites rapidly increases as stellar mass decreases \citep{2015Wetzel}.\ These works suggest an environmental and mass dependence on the quenching mechanisms of dwarf satellite galaxies.\ Higher-mass satellites are thought to be less susceptible to the effects of their hosts and hold on to their HI reservoirs, allowing them to continue forming stars throughout their evolution \citep[e.g.][]{Simpson2018,2019garrisonkimmel}.\ On the other hand, the weaker gravitational potentials of lower-mass satellites make them more susceptible to both internal (i.e.\ star-formation feedback) and external (i.e.\ ram-pressure and tidal stripping) processes.\ At the lowest stellar masses, ultra-faint dwarfs may have been quenched via reionization regardless of their environment \citep{2000Bullock,2014Brown,2017Jeon,2021Applebaum,Sand2022}.\ Combining these results from observations with modern hydrodynamical simulations provides useful insight into the evolutionary processes that occur at the smallest mass scales.\ 

There has been general consistency between observations of the LG and recent simulations of LG and MW-like systems \citep[e.g.][]{2016Fattahi,2016Sawala,Wetzel2016,Simpson2018,2019garrisonkimmel,2020Libeskind,2021Engler,2021Font,2021Applebaum,Akins2021,2022Samuel}, including investigations into the mass and time scales for satellite quenching.\ However, there still exist some tensions as we begin to explore larger samples of satellite systems beyond the LG \citep{2022Sales}. One such tension is the contrast between both the quenched fractions of the LG and simulations of LG- and MW-like systems with those in the Satellites Around Galactic Analogs survey \citep[SAGA;][]{GehaSAGA,2021Mao}.\ 

The SAGA survey aims to identify, spectroscopically confirm, and characterize the satellite population of nearby MW-like hosts down to satellites as faint as Leo I ($M_r=-12.3$ mag).\ The H$\alpha$ emission from the single-fibre spectroscopic follow-up of the SAGA satellites was used to characterize them as quenched or star-forming with the vast majority falling into the latter category.\ Even so, these single-fibre spectroscopic observations may have underestimated the number of star-forming satellites, which motivated \citet{2021Karunakaran} to use archival \textit{GALEX} UV imaging as an alternative spatially resolved star-formation tracer.\ Their findings increased the fraction of star-forming satellites in the second stage release of the SAGA survey (hereafter, SAGA--II; \citealt{2021Mao}) from $\sim$85\% to $\sim$95\%.\ In addition to this re-characterization of the star-forming nature of these systems, they provided a comparison of the fraction and number of quenched satellites in the APOSTLE \citep[A Project Of Simulating The Local Environment;][]{2016Fattahi,2016Sawala} and Auriga \citep{Simpson2018} hydrodynamical simulation suites.\ This comparison used similar host sample sizes and applied similar stellar mass cuts to demonstrate a potential tension between the surplus of star-forming satellites in observations and the relative dearth in simulations (see also \citealt{Simpson2018, Akins2021, 2022Samuel, 2023Engler}), even when incompleteness corrections are accounted for.\ 

Given that many simulations use the Local Group as a useful benchmark, it is interesting to ask: Are the Local Group and its quenched fraction typical? Also, extending the study of satellite systems and their quenched fraction to lower luminosities and masses than SAGA--II is an essential next step for understanding what drives the evolution of the smallest galaxies.\ These lower-mass galaxies are even more susceptible to the environmental effects of their hosts.\ Understanding whether the trends we see in the lower mass regime within the Local Group are present beyond it can help constrain whether or not the Local Group is typical.\

This tension in the quenched fraction between SAGA--II and both the LG and simulations may lend some credence to the latter two being less representative of the broader MW-like population.\ However, \citet{2022Font} have presented a potential solution to this tension.\ By applying the SAGA--II absolute magnitude cut to the satellites around MW-like hosts in the ARTEMIS (Assembly of high-ResoluTion Eagle-simulations of MIlky Way-type galaxieS) simulation suite \citep{2020Font}, they showed that the quenched fraction can be decreased from $\sim95\%$ to $\sim75\%$, in their lowest stellar mass bin, $\mathrm{log}(M_*/M_{\odot})\sim6.8$.\ However, in their earlier work they had estimated a by-eye surface brightness limit for the SAGA--II survey of $\mu_{\mathrm{eff},r}\sim25\mathrm{mag\,arcsec^{-2}}$ \citep{2021Font}.\ This estimate is consistent with the empirical completeness limit derived by \citet{2021kadofong} for the Dark Energy Survey low surface brightness (LSB) catalog \citep[][]{2021Tanoglidis}.\ \citet[][]{2021Mao} had used the Dark Energy Survey LSB catalog to cross-match with their photometric catalog to demonstrate that they are not missing any LSB satellite candidates.\ Therefore, the agreement between the limits from \citet{2021Font} and \citet{2021kadofong} suggests that the SAGA--II photometric catalogs may be missing LSB satellite candidates.\ Combining this proposed surface brightness limit, in addition to an absolute magnitude limit, \citet{2022Font} found a significant, $\sim$50\%, reduction in the quenched fraction in their lowest stellar mass bin relative to the total ARTEMIS satellite sample.\ This by-eye surface brightness limit is a potentially useful metric to investigate further with other simulated and observational samples, particularly those that are complete towards lower surface brightness \citep[e.g.][]{2022Zaritsky}.\ 

There have been a plethora of observational efforts to characterize the satellite populations in the nearby Universe \citep[e.g.][]{Chiboucas2009,Chiboucas2013,2014MerritLSBsM101,Sand2014,2015Karachentsev,Sand2015,Carlin16,PISCeS-CenAsats, JavanmardiDGSAT, Bennet2017, 2017Muller, Smercina2018, Bennet2019, CarlstenSBFM101, Crnojevic2019, 2020Bennet, 2021MutluPakdil, 2022MutluPakdil, 2022Bell}.\ A complementary survey to SAGA--II in the Local Volume is the Exploration of the Local VolumE Satellites (ELVES) survey \citep{ELVESI}.\ This survey of nearby ($<12$ Mpc) massive hosts builds upon earlier work within the Local Volume.\ An advantage of the ELVES survey is its ability to reach fainter systems and estimate robust surface brightness completeness limits as a result of performing their own photometry.\ On the other hand, their confirmation efforts via surface brightness fluctuation-based distance measurements are susceptible to systematic effects and may be less robust relative to the SAGA--II spectroscopic follow-up.\ For instance, some satellite candidates from ELVES have been shown to be at different distances than that inferred from surface brightness fluctuations \citep[e.g.][]{2020Karunakarana,2022Karunakaran}.\ Irrespective of these differences, having another survey to compare to the LG is critical.\ \citet{ELVESI} find broad agreement with the observed quenched fractions in the LG and do not agree with the low quenched fractions from the SAGA--II satellites even when incompleteness corrections are taken into account.\ The ELVES sample provides an interesting opportunity to study the quenched fraction and number of satellites down to $M_V\sim-9$.\ Its relatively high completeness also allows us to test the effect of the SAGA absolute magnitude cut (i.e.\ $M_r=-12.3$) as well as surface brightness effects.\

In this paper, we investigate the number of star-forming satellites and quenched fractions in the ELVES and SAGA--II satellite samples and explore the role different selection criteria (e.g. magnitude and surface brightness limits) play in measuring these quantities.\ We hope to better understand the Local Group in the broader context of satellite systems as well as whether or not the larger statistical SAGA--II sample suffers from systematic biases that prevent drawing definitive conclusions about satellite populations beyond the Local Group.\ In Section \ref{sec:sample}, we provide an overview of the SAGA--II and ELVES satellite samples.\ We describe two different criteria for selecting `star-forming' dwarfs in Section \ref{sec:defQSF} and then move on to discuss the various selection effects that we explore in Section \ref{sec:seleffects}.\ We briefly summarize the implication of our results in Section \ref{sec:summary}.

\section{Satellite Samples}\label{sec:sample}
\subsection{The Satellites Around Galactic Analogs (SAGA) Survey}
The SAGA--II sample \citep[][hereafter, \citetalias{2021Mao}]{2021Mao} consists of 127 satellites around 36 nearby (25-40 Mpc) Milky Way-like systems.\ The primary selection criteria of these hosts is their $K$-band luminosity $(-23>M_K>-24.6)$.\ The halo masses of these systems $(M_{halo}\sim(0.7-2)\times10^{12}\msun)$ were estimated to also cover the halo mass range encapsulating the MW's halo mass \citep[\citetalias{2021Mao}, see also][]{2020Nadler}.\ Most host systems are in relative isolation (i.e.\ field-like environments), while a few are analogs of the LG (see \citetalias{2021Mao} for details).\

The satellites in SAGA--II were initially selected from imaging catalogs (i.e.\ Legacy Survey, SDSS, and DES).\ Satellite candidates without extant redshift measurements were observed through various optical spectroscopic campaigns.\ Overall, 80\% of satellite candidates with $M_r\leq-12.3$ (assuming the host's distance) were spectroscopically targeted.\ The entire SAGA--II sample contains a total of 127 satellites that project within 300 kpc with a recessional velocity within $275\,\kms$ of their putative host.\ To avoid high levels of background light, satellites with projected distances less than 15 kpc are excluded for spectroscopic follow-up \citep{ELVESI}.\ Furthermore, there are four satellites that fall below the aforementioned SAGA--II absolute magnitude limit.\ While we include these four satellites in our analysis, we note that the results presented here remain unchanged if they were to be removed.

As previously mentioned, a large number of satellites in SAGA--II are actively star-forming as determined by their H$\alpha$ measurements.\ In a subsequent study, \citet{2021Karunakaran} presented UV measurements leveraging the spatial coverage of archival \textit{GALEX} imaging to show that an even larger number of the SAGA--II satellites have undergone recent star formation.\ The apparent dearth of quenched satellites in these systems, relative to the Local Group, points toward potential survey incompleteness.\ Alternatively, the larger SAGA--II sample may be more representative of Milky Way-mass systems, making the Local Group an outlier.\

In order to characterize their survey incompleteness and potential interloper fraction, \citetalias{2021Mao} perform extensive modeling of their confirmed and unconfirmed targets in conjunction with mock sample selection from dark matter-only simulations.\ In particular, they provide incompleteness/interloper-corrected quenched fraction estimates based on the detection of H$\alpha$, which we adopt here.\ These corrections assume that all of the potential satellites that were missed in the SAGA--II spectroscopic follow-up are quenched.\

We adopt the photometric and derived properties (e.g.\ magnitudes, surface brightness, stellar masses, etc.) from SAGA--II.\ We adopt the UV-derived star-forming quantities from \citet{2021Karunakaran} for the SAGA--II sample.\ The spatial coverage of the \textit{GALEX} UV imaging ensures that we capture any star formation-related emission across these satellites which may be missed by the smaller optical fibers.\

\subsection{The Exploration of Local VolumE Satellites (ELVES) Survey}
The ELVES survey \citep[][hereafter, \citetalias{ELVESI}]{ELVESI} builds upon earlier surveys of Local Volume (D$<$12 Mpc) systems, primarily that of \citet{Carlsten2020}.\ The ELVES sample consists of 31 Local Volume hosts whose primary selection criteria is their $K-$band luminosity, $M_K<-22.1$.\ The satellites around all hosts have been cataloged within 150 kpc, however, most have been surveyed within 300 kpc.\ It should be noted that the satellite systems for 5 ELVES hosts (MW, M31, Cen A, NGC5236, and M81) are taken from the literature \citep{2012McConnachie,Chiboucas2013,2015Muller,2017Muller,2018Muller,Crnojevic2019}.\

The ELVES survey uses a combination of archival CFHT/MegaCam \citep[i.e.][]{Carlsten2020} and DECaLS imaging primarily for source detection.\ The confirmation of the association of these satellites is through surface brightness fluctuation (SBF) distance estimates.\ The SBF method, tuned for LSB dwarfs \citep[e.g.][]{2019Carlsten}, is employed on deeper Hyper Suprime-Cam, Gemini, and Magellan imaging.\ We note that there may always be cases where the SBF method faces difficulty, particularly for faint blue, clumpy/irregular (i.e.\ star-forming) systems \citep[e.g.][]{2019Carlsten,2021Greco,2020Karunakarana,2022Karunakaran}.\ These SBF distances are used to classify a satellite as ``confirmed'', ``background'', or ``unconfirmed/candidate''.\ The total ELVES sample consists of 338 confirmed satellites across 30 hosts\footnote{The lone host yet to be analyzed is NGC3621} and 106 additional satellite candidates with forthcoming distance estimates.\ While these additional satellites will eventually provide useful statistical confidence, we only include ``confirmed'' satellites in this work.

Due to the excellent quality of the data, the ELVES sample is capable of reaching faint satellite luminosities.\ Through robust artificial galaxy injection and recovery simulations, \citetalias{ELVESI} determined the ELVES sample is complete down to $M_V\sim-9$ and $\mu_{0,V}\sim26.5\,\mathrm{mag\,arcsec^{-2}}$.\ In addition to deriving these completeness limits, \citetalias[][]{ELVESI} perform analogous modeling to \citetalias{2021Mao} to determine how many of the 106 satellite candidates without distances are real satellites, finding that as many as $\sim50$ out of 106 may be real.\

Another crucial difference between the ELVES and SAGA--II surveys is their survey coverage.\ While all of the SAGA--II systems have photometric catalogs within a projected 300 kpc radius of their hosts, 21 out of 30 ELVES hosts have coverage out to this projected radius with the remaining having coverage to at least 150 kpc.\ We only consider ELVES satellites with projected distances beyond 15 kpc, to match the SAGA--II cut.\ This removes 6 out of 338 confirmed ELVES satellites.

We adopt all of the photometric properties, physical quantities, and classifications derived for the hosts and satellites in the ELVES sample.\ We compute $r-$band ($i-$band) apparent magnitudes of any ELVES satellite with only $g-$ and $i-$band ($r-$band) photometry using the relations of \citet{Carlsten2021} in order to conduct seamless comparisons to SAGA--II satellites.\ Additionally, \citetalias{ELVESI} present UV photometry for 271 satellites with available \textit{GALEX} imaging.\ We also adopt these UV measurements and derive any distance-dependent quantities (i.e.\ star-formation rates) assuming a satellite's host distance.\

\section{Identifying Quenched and Star-forming Satellites}\label{sec:defQSF}
Classifying a galaxy as quenched or star-forming can be done through a variety of methods.\ Here, we outline two possible methods of separating quenched and star-forming satellites in the ELVES and SAGA-II samples and use these as our star-forming definitions.

The first method is a photometric selection established by \citet{Carlsten2021} where they separated quenched, early-type galaxies (ETGs) and star-forming, late-type galaxies (LTGs) with the relation $g-i=-0.067\times M_V - 0.23$ in the colour--absolute $V-$band magnitude plane.\ We use this relation for both the ELVES and SAGA--II samples in Section \ref{sec:seleffects} to separate quenched and star-forming satellites.\ We convert any $g-r$ colours to $g-i$ using the conversion provided in \citet{Carlsten2021}.\

\begin{figure}
	\includegraphics[width=\columnwidth]{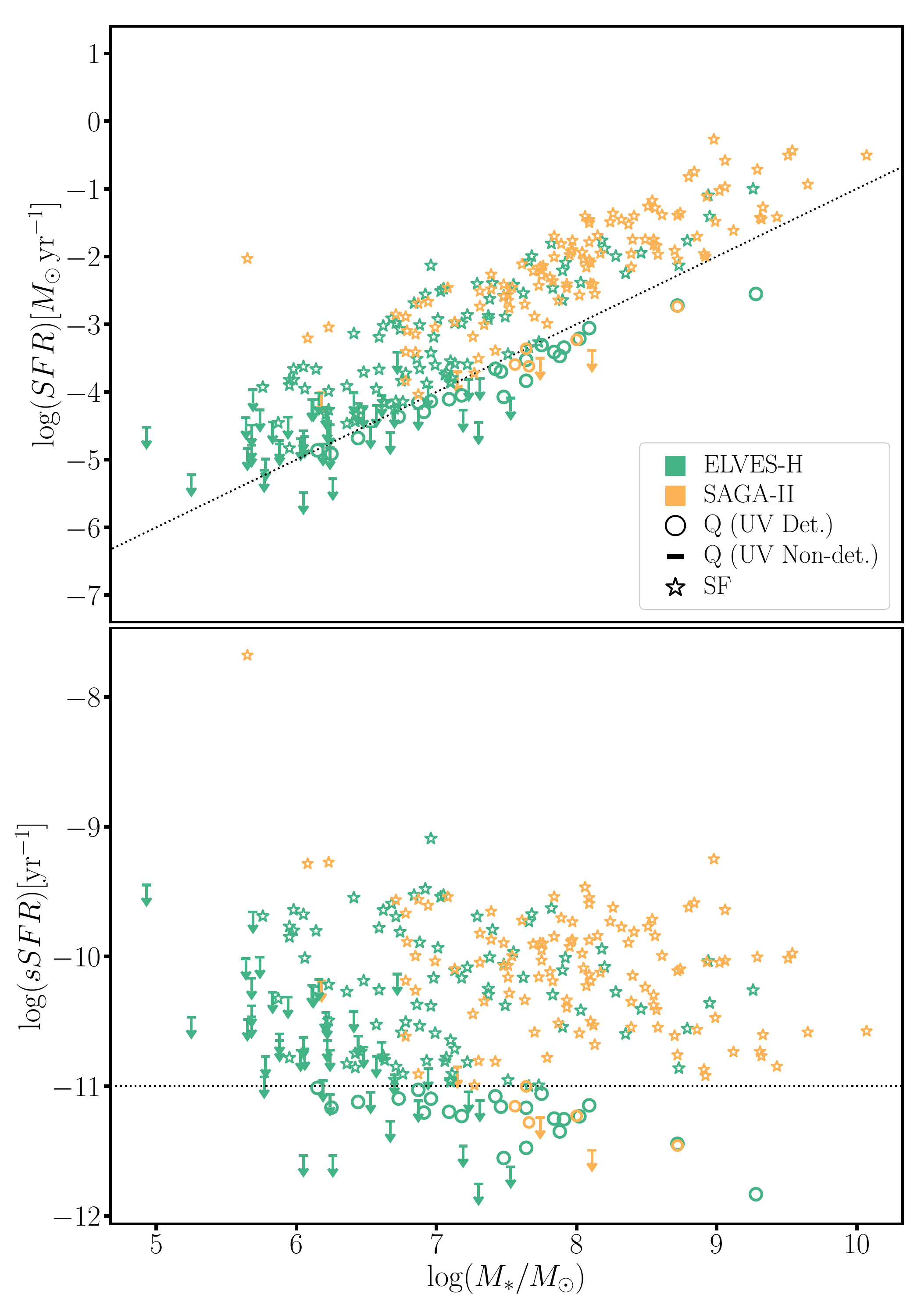}
    \caption{A comparison of SFR (top) and sSFR (bottom) as a function of stellar mass for the ELVES (green) and SAGA-II (orange) satellite samples.\ The ELVES sample shown in this figure includes only satellites with available \textit{GALEX} UV data and with MW-like or less massive hosts (see Section \ref{sec:seleffects} for more detail).\ Satellites that are detected in UV and have log(sSFR$[\mathrm{yr^{-1}}]$)$\geq-11$ are star-forming and shown as stars.\ Quenched (indicated as Q in the legend) satellites that are detected in UV but do not meet the aforementioned sSFR threshold are shown as open circles, while those that are not detected in UV are upper limits and shown as bars with downward arrows.\ The dotted line in both panels show a constant log(sSFR$[\mathrm{yr^{-1}}]$)$=-11$.}
    \label{fig:sfrcomp}
\end{figure}

The second method is via UV emission, particularly the specific star formation rate (sSFR), indicating relatively recent (i.e.\ $\sim200$ Myr) star formation.\ We use here the NUV photometry derived for the SAGA--II sample from \citet{2021Karunakaran} and for the ELVES sample from \citetalias{ELVESI}.\ \citet{2021Karunakaran} used a simple star-forming criterion, i.e.\ $(S/N)_{NUV}>2$, for the SAGA--II satellites and they found a high correspondence of satellites with both H$\alpha$ and NUV emission.\ While this simple definition may suffice for more distant star-forming systems, NUV imaging of more nearby quenched systems may be sensitive to NUV emission from relatively older stellar populations \citep[][]{2009Lee}.\ 

To circumvent this potential bias, we derive NUV star-formation rates using the SFR relation from \citet{2006IglesiasParamo}.\ These SFRs have not been corrected for internal infrared extinction which is likely significant for more massive satellites and likely negligible for low-mass ones \citep[][]{2015Mcquinn}.\ We then determine an sSFR by taking the ratio of the NUV-derived SFRs and stellar masses from the ELVES and SAGA--II sample catalogs.\ In Figure \ref{fig:sfrcomp}, we show these derived quantities as a function of stellar mass for both the ELVES (green) and SAGA--II (orange) samples\footnote{As noted in \citet{2021Karunakaran}, LS-330948-4542, the lowest mass SAGA--II satellite ($\mathrm{log}(M_*/M_{\odot})\sim5.6$), appears to have a size that is severely underestimated in the SAGA--II catalog which leads to a subsequent erroneous calculation of its stellar mass.\ Nonetheless, we make no attempt to correct the reported properties of this object in the current study.}.\ Satellites that are detected in UV and with $\mathrm{log}(sSFR)\geq-11$ are considered star-forming, while those that fall below this limit or are undetected, i.e.\ $(S/N)_{NUV}<2$, in NUV are quenched.\ Quenched satellites are shown as circles and NUV non-detections are shown as bars with downward arrows, while star-forming satellites are represented by stars.\ The ELVES sample probes lower stellar masses and most of these lower mass satellites are classified as quenched.\

We note that there are minor differences in the ELVES satellite sample sizes as only a subset (271/338) have available UV imaging.\ Furthermore, in cases where ELVES photometry is missing in a particular band, we assume the average $g-r$ or $g-i$ colour of the sample to determine the required quantity.\ Three relatively bright ELVES satellites (NGC4656, NGC247, IC239) and the Milky Way satellites included in ELVES have neither $g,\,r,\,i\,$ nor UV photometry.\ We substitute their morphological classifications (i.e.\ ETG to quenched and LTG to star-forming) in our comparisons given that the three ELVES satellites are more luminous systems and the Milky Way satellites are typically less ambiguous in terms of their state of star formation.\ Indeed, the ETG/quenched Milky Way satellites are all undetected in HI and have very low HI mass to V-band luminosity ratio upper limits, $-6\lesssim \mathrm{log}(\frac{\mhi}{\LV}[\frac{\msun}{\lsun}]) < -3$ \citep{Putman2021}.

\section{Selection Effects on the Quenched Satellite Population}\label{sec:seleffects}

\subsection{Host and Photometric Selection Criteria}
There are clear differences in the host and satellite properties between the ELVES and SAGA-II samples.\ We briefly describe selection criteria in order to have a fair comparison of these two samples.\

We first address the differences in hosts.\ As noted by \citetalias[][]{ELVESI}, there are several host galaxies in the ELVES sample that would not meet the SAGA--II host criteria: luminosity (i.e.\ $M_K<-24.6$), isolation (i.e.\ nearby bright companion), and/or halo mass (i.e.\ $M_{\mathrm{halo,group}}<10^{13}M_{\odot}$).\ We remove eight hosts (too bright: M104 and M31; nearby massive companion: NGC 1808, NGC 5194, NGC 3379, and M81; high halo mass: Cen A and NGC 3627) listed in \citetalias[][]{ELVESI} that fail any of these criteria, see their Section 7.1 for more details.\ In addition to the aforementioned eight hosts, we find the inclusion or exclusion of six ELVES hosts (NGC 628, NGC 3344, NGC 3556, NGC 4517, NGC 4631, and NGC 4736) that fall below the SAGA--II $M_K$ limit (i.e.\ $M_K>-23$) make marginal changes to our results.\ Similar to the conclusions of \citetalias[][]{ELVESI}, we opt to keep these systems for the purpose of sample size.\

We now turn to a couple of photometric selection criteria that can affect the comparison between these two samples.\ The first is the absolute magnitude limit of the SAGA--II sample, $M_r\leq-12.3$, which will apply to the ELVES sample as well.\ We note that there are other photometric cuts applied within the SAGA--II survey with respect to spectroscopic follow-up and discuss their relevance below.\ The second photometric criterion we consider in addition to the $M_r$ cut is a surface brightness cut of $\mu_{\mathrm{eff},r} <25\,\mathrm{mag\,arcsec^{-2}}$.\ This selection cut was recently suggested by \citet{2022Font} to reconcile differences between the SAGA--II sample and hydrodynamical simulations.\ Here, we apply a similar cut in surface brightness to see its resulting effects on the number and fraction of quenched satellites in the ELVES sample.\

We compute the mean surface brightness values for the ELVES sample in the same manner as SAGA--II (see their Eq.\ 2) by using the previously computed $r-$band magnitudes (see Section \ref{sec:sample}) and determining the apparent effective radii using the physical radii and host distances tabulated in \citetalias{ELVESI}.\ For the Milky Way satellites with only $V-$band photometry, we approximate their mean $r-$band surface brightness values assuming $V-r\sim0.4$ \citep[][]{Crnojevic2019} and an exponential light profile to move from central to effective values, $\mu_{\mathrm{eff}}-\mu_0\sim1.123$ \citep[][ see also \citealt{2005Graham}]{2022Font}.\ Given the various selection criteria described and the different star-forming definitions, we have provided a summary of the number of satellites that remain after each cut in Table \ref{tab:selection}.

\begin{table*}
    \begin{center}
    \caption{Breakdown of the total number, N, of ELVES Satellites and mean quenched fractions, $\langle$QF$\rangle$, for each Selection Criteria and Star-forming Definition.\ In addition, we list these values when only considering satellites that project within 150 kpc ($\mathrm{N_{150}}$ and $\langle$QF$\rangle_{150}$).\ The SAGA--II values are listed at the bottom for reference.}
    \begin{tabular}{lcccccccc}
        \hline
        Selection & \multicolumn{4}{c}{Colour-Magnitude} & \multicolumn{4}{c}{UV+sSFR}\\
       Criteria & N & $\langle$QF$\rangle$ & $\mathrm{N}_{150}$ & $\langle$QF$\rangle_{150}$ & N & $\langle$QF$\rangle$ & $\mathrm{N}_{150}$ & $\langle$QF$\rangle_{150}$\\
        \hline
        Total & 338 & 72\% & 214 & 75\% & 271 & 58\% & 183 & 62\% \\
        ELVES-H & 181 & 66\%  & 121 &70\% & 168 & 50\%  & 113 & 55\% \\
        ELVES-M & 98 & 37\%  & 65 & 42\% & 89 & 27\%  & 59 & 28\% \\
        ELVES-SB & 82 & 34\%  & 54 & 39\% & 75 & 27\%  & 49 & 30\%\\
        \hline
        SAGA--II & 127 & 11\%  & 72 & 16\% & 127 & 9\%  & 72 & 15\%\\
        \hline
    \end{tabular}
    
    \label{tab:selection}
    \end{center}
\end{table*}

\begin{figure*}
    \includegraphics[width=\textwidth]{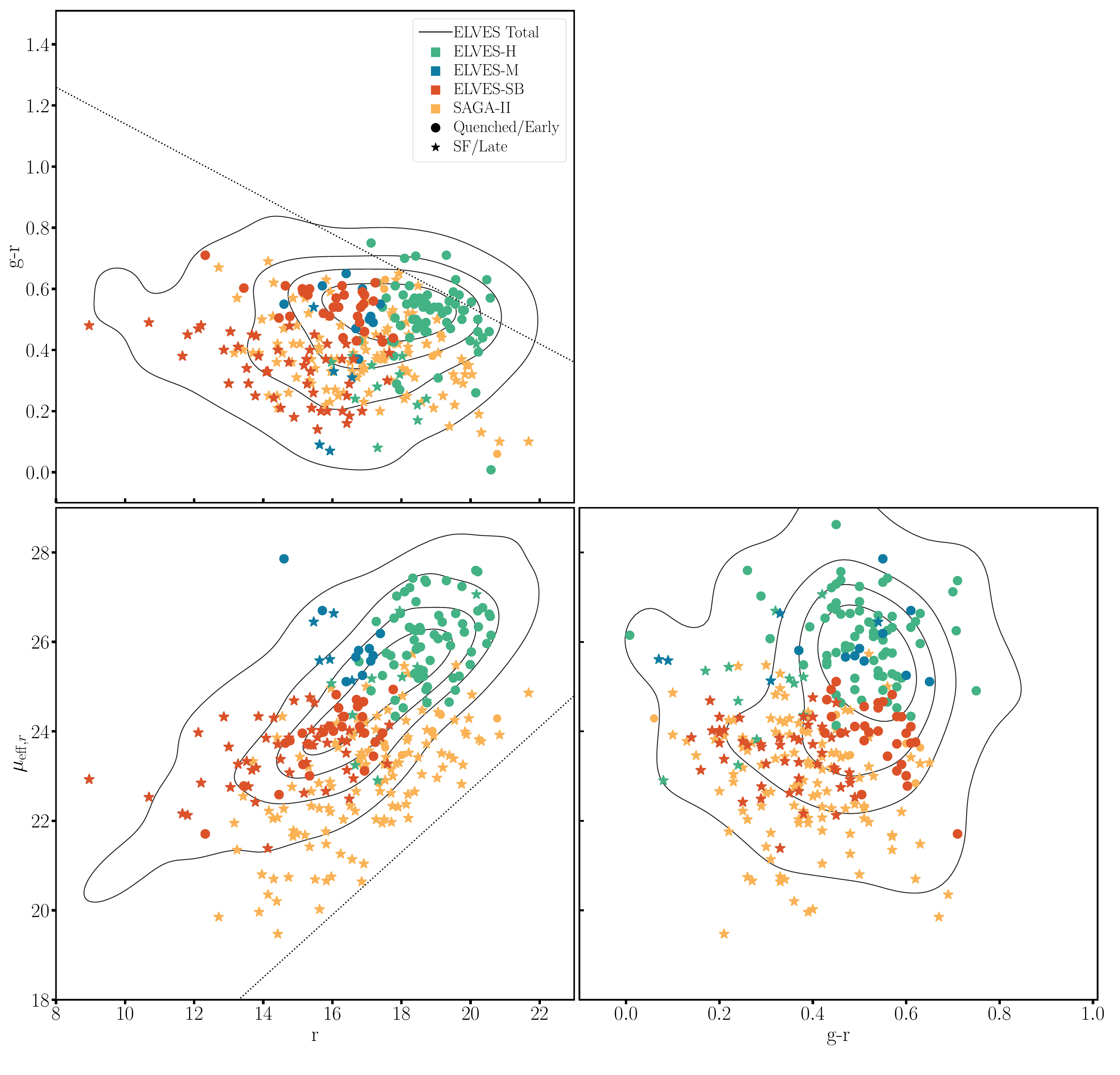}
    \caption{A comparison of the apparent photometric properties of the ELVES and SAGA-II samples.\ g--r vs r--band apparent magnitude (top-left), mean r--band surface brightness at the effective radius ($\mu_{\mathrm{eff},r}$) vs r--band apparent magnitude (bottom-left), and $\mu_{\mathrm{eff},r}$ vs g--r (bottom-right).\ The ELVES sample is subdivided by the various observational selection criteria as detailed in Section \ref{sec:seleffects} and summarized in Table \ref{tab:selection}.\ For reference, the contours show the KDE distribution of all satellites classified as ``Confirmed'' in the ELVES sample.\ The SAGA--II sample is shown in orange.\ Additionally, we include the relations defining the ``primary spectroscopic targeting regions'' from the SAGA--II survey in the top- and bottom-left panels as the diagonal dotted lines.\ The quenched and star-forming definitions are defined as in Fig.\ \ref{fig:sfrcomp}. }
    \label{fig:photpropcomp}
\end{figure*}

In Figure \ref{fig:photpropcomp}, we show a comparison of three essential photometric properties of these satellites: $r-$band apparent magnitude, surface brightness ($\mu_{\mathrm{eff},r}$), and colour ($g-r$).\ The total ELVES confirmed satellite sample with none of the above cuts is represented by the contours in each of these planes.\ The compounding selection cuts on the ELVES sample are shown in different colours: the host cut is shown in green (ELVES-H), the host+SAGA--II absolute magnitude cut in blue (ELVES-M), and the host+SAGA--II absolute magnitude+surface brightness cut in red (ELVES-SB).\ It should be noted that since these cuts are compounding, the sub-samples are included in their parent sample, i.e.\ all red symbols are included in the same sample as the blue and green symbols.\

We can see from this figure that there is a clear separation between the ELVES sample with only the host cut (green+blue+red) and the two sub-samples (blue and red).\ For reference, we also include the relations from \citetalias{2021Mao} that are used to define the primary targeting region for their spectroscopic follow-up (see their Eqs.\ 3a and 3b) as dotted lines in the upper- and lower-left panels.\ In the upper-left panel, a small number of satellites do not satisfy the colour-magnitude relation (diagonal dotted line) that SAGA--II uses to target candidates spectroscopically.\ Since these satellites do not satisfy the SAGA--II absolute magnitude cut, they will not skew our comparisons to the SAGA--II sample.\ In the surface brightness-magnitude plane, all ELVES satellites would qualify for spectroscopic follow-up as they lie above the SAGA--II relation.\ The minimum angular size imposed on ELVES sources (circular radius $\sim4''$) is also clear relative to the SAGA--II sample, however, this should have minimal impact on the interpretation given the difference in distances to systems in each survey.\ The star-forming populations in both the ELVES-SB and SAGA--II samples reside in similar photometric parameter space.\ While there exist some high surface brightness, bright ELVES satellites that are analogs to those in the SAGA--II sample, they are members of host systems that do not satisfy the host criteria.\

Within the ELVES sample, the vast majority of satellites that are removed by the SAGA--II absolute magnitude cut alone are quenched, faint, relatively red, and fall towards fainter effective surface brightness.\ We can also see that there is only a marginal difference between the ELVES-M and ELVES-SB sub-samples.\ In this case, the absolute magnitude cut effectively removes the LSB satellites and negates any effect from the surface brightness cut.\ We explore the implications of this in the following subsection.\

\subsection{Quenched Fractions and Satellite Counts}
With both the quenched/star-forming definitions and selection effects established above, we explore their effects on the quenched fractions and the number of satellites per host in the ELVES and SAGA--II satellites samples.\ We first consider all ELVES hosts that meet the aforementioned host criteria and then shift our focus to the inner 150 kpc, the maximum radius out to which all ELVES hosts have been completely surveyed.\ In addition to this latter point, we briefly explore the quenched fractions and satellite counts as a function of radius.\

\begin{figure*}
	\includegraphics[width=\textwidth]{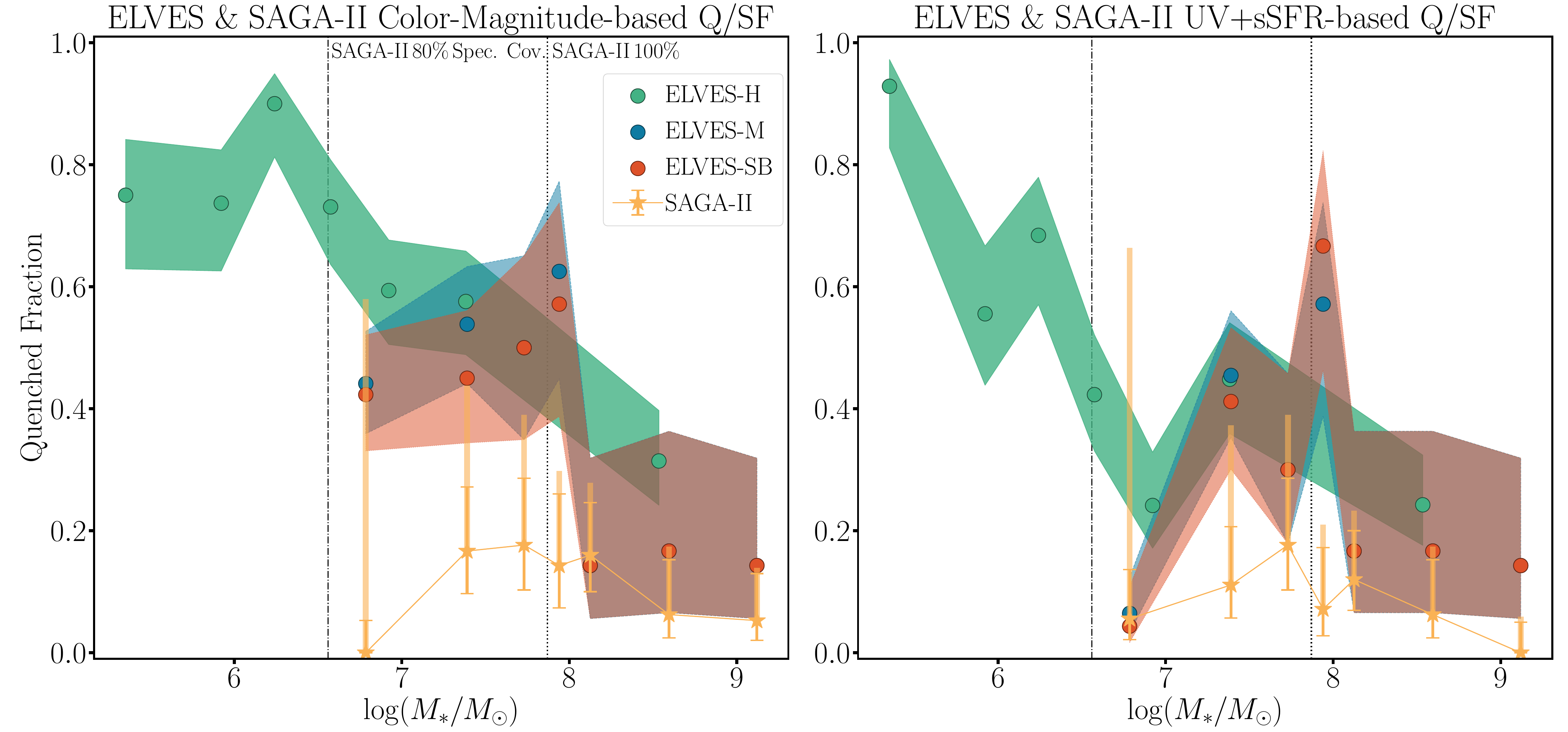}
    \caption{Satellite quenched fraction as a function of stellar mass.\ The ELVES sample is separated into different subsets by colour in order to highlight the effect of compounding selection criteria (see Section \ref{sec:seleffects} and Table \ref{tab:selection}).\ The definition of a quenched satellite is different in each panel: left--quenched and star-forming satellites are divided by a relation in the (g-i)--M$_V$ plane, right--quenched satellites in both samples are either NUV non-detections or their $\mathrm{log(sSFR)}<-11$.\ The shaded regions and error bars show the 68\% confidence intervals and the extended bars from the symbols show the confidence intervals plus the incompleteness/interloper corrections.\ The vertical dash-dotted and dotted lines show the SAGA--II 80\% and 100\% spectroscopic coverage limits.
    }
    \label{fig:qf}
\end{figure*}

In Figure \ref{fig:qf}, we show the ELVES and SAGA--II quenched fractions as a function of satellite stellar mass.\ Each panel uses one of the quenched/star-forming definitions established in Section \ref{sec:defQSF}.\ The binning of the SAGA--II, ELVES-M, and ELVES-SB samples attempts to follow the original binning in \citetalias{2021Mao} in the same manner as in \citet{2021Karunakaran}.\ Since the ELVES-H sample spans the widest range in stellar mass, we have binned this sample to roughly equate the number of total satellites (i.e.\ the denominator of the quenched fraction) in each stellar mass bin.\ In the left panel, we use the colour-magnitude relation from \citet{Carlsten2021} to separate quenched and star-forming satellites for both satellite samples.\ The right panel employs the NUV detection and sSFR threshold to distinguish star-forming and quenched satellites, again, for both samples.\ Within each of these panels, we compute the quenched fractions after applying the three compounding selection criteria described above and follow the same colour scheme as in Figure \ref{fig:photpropcomp}.\ The shaded regions (ELVES) and error bars (SAGA--II) are the 68\% confidence intervals computed using the Wilson score interval \citep{brown2001}.\ The extended lighter-coloured bars from the SAGA--II quenched fractions show the incompleteness/interloper corrections estimated by  \citetalias{2021Mao}.\ The dash-dotted and dotted lines, respectively, show the SAGA--II stellar masses at which 80\% and 100\% of satellites have been targetted spectroscopically\footnote{Following \citet{2021Karunakaran}, we convert the $M_r=-12.3$ and $M_r=-15.5$ limits to $M_*$ using the stellar mass conversion relation from  \citetalias{2021Mao} and the average SAGA--II colour, $g-r\sim0.39$.}.\

In the left panel, we can see that the application of these additional selection criteria (ELVES-M and ELVES-SB) does marginally decrease the estimated quenched fractions and bring them into better agreement with the SAGA--II sample.\ Shifting to the right panel, which shows the UV+sSFR-based quenched/star-forming classification, we see a decrease in the ELVES quenched fractions below $\mathrm{log}(M_*/M_{\odot}) \sim8$.\ This shift brings the ELVES sample into stronger agreement with the SAGA--II than the colour-magnitude (or morphology) definition(s).\ Although the partial UV coverage for the ELVES sample does affect the number of satellites in each subset, there are only marginal ($<10\%$) differences between the ELVES-M and ELVES-SB in either of the quenched/star-forming panels.\ It is interesting to note the general agreement between the ELVES-H and SAGA--II sample in the right panel, particularly in the $\mathrm{log}(M_*/M_{\odot}) \sim6.8$ bin.\ While the additional selection criteria (ELVES-M and ELVES-SB) have an effect on the quenched fraction, this agreement between the ELVES-H and SAGA--II values may also stem from several compounding factors such as the aforementioned UV coverage, choice of binning, inclusion of lower mass hosts, etc.\ We note that the spike in the quenched fraction in the $\mathrm{log}(M_*/M_{\odot}) \sim8$ bin is a result of adopting the SAGA--II stellar mass bins.\ Another factor to consider is the SFR relation we have used in this work from \citet{2006IglesiasParamo}.\ As noted by \citet{2023Greene}, it is possible that we are overestimating our quenched fractions since these relations are calibrated to higher metallicity systems.\ However, we find commensurate SFR estimates for SAGA satellites from our NUV analysis and forthcoming H$\alpha$ imaging (Jones et al., in prep), suggesting that a better understanding of SFR relations is required, particularly for low mass galaxies that are not undergoing large star forming episodes.\ Nevertheless, we adopt this latter star-forming definition as our preferred definition as it is physically motivated (i.e.\ sensitive to significant recent star-formation) and is widely adopted in both observations \citep[e.g.][]{2013Karachentsev,2013Wetzel} and simulations \citep[e.g.][]{2019Pallero,Akins2021,2021Joshi}.\ We keep the aforementioned caveats in mind and consider these two star-forming definitions as upper and lower bounds on the ``true'' quenched fractions.\ We provide the mean quenched fractions, $\langle$QF$\rangle$, for each of these selection criteria and star-forming definitions in Table \ref{tab:selection}.

\begin{figure*}
	\includegraphics[width=\textwidth]{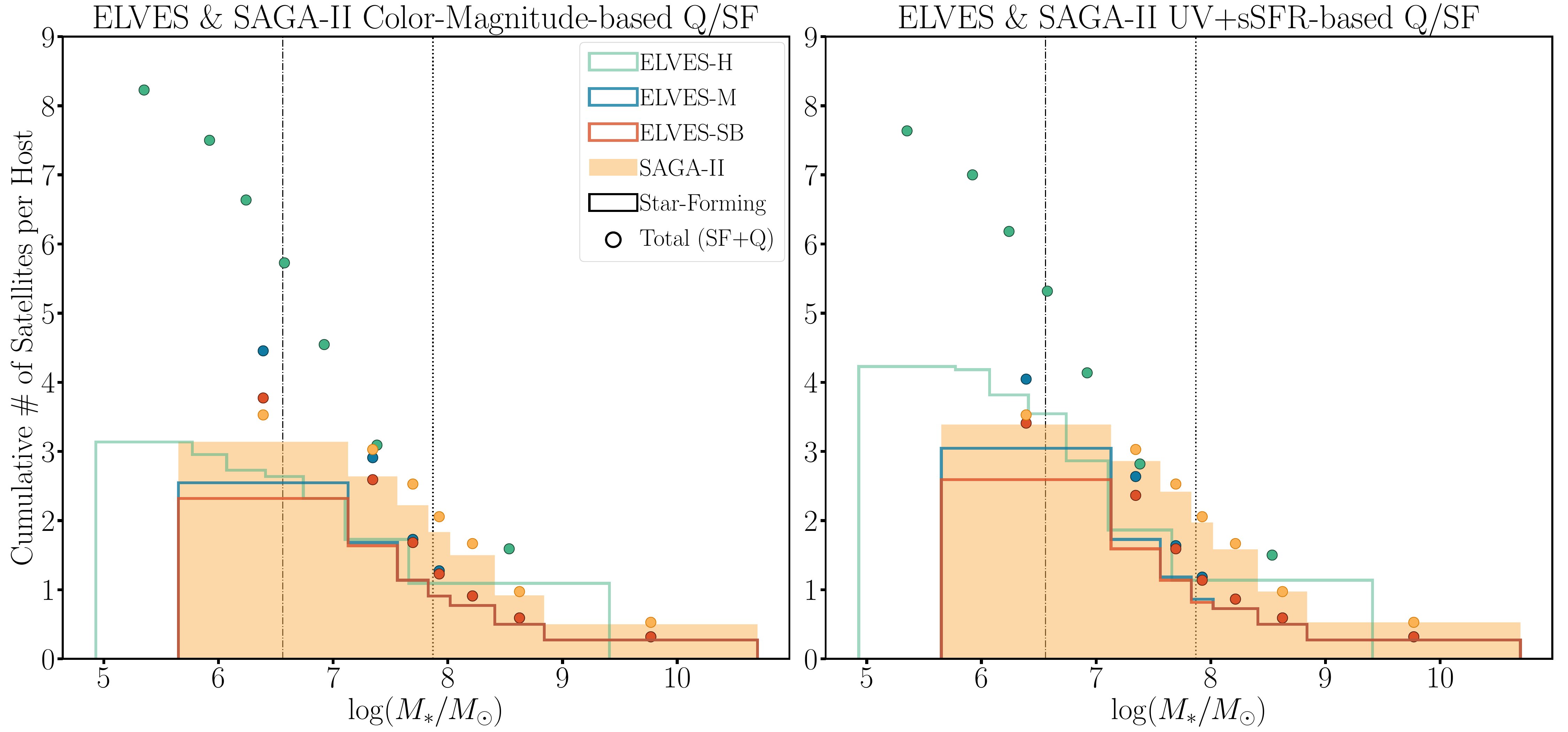}
    \caption{Cumulative number of total (filled circles) and star-forming (open and filled histograms) satellites for the various ELVES sample subsets in comparison to the SAGA--II sample.\ The subsets are coloured as in Figures \ref{fig:photpropcomp} and \ref{fig:qf}.\ }
    \label{fig:chist}
\end{figure*}

In both of the panels in Figure \ref{fig:qf}, we can see the minimal difference the additional surface brightness criterion (ELVES-M vs ELVES-SB) makes to the resulting quenched fractions.\ To better understand this, we investigate the number counts of satellites in the ELVES and SAGA--II samples.\ Figure \ref{fig:chist} shows the cumulative number of the star-forming (open and filled histograms) and total (i.e.\ star-forming+quenched; filled circles) satellites per host as a function of stellar mass.\ Both panels show the same quenched/star-forming classification methods as in Figure \ref{fig:qf}.\ The tight correspondence between the star-forming and the total number of SAGA--II satellites hint at a potential dearth of quenched satellites.\ It is evident that the bulk of the SAGA--II satellites build up at intermediate to high stellar masses.\ On the other hand, the ELVES star-forming satellites more slowly increase up until the final two stellar mass bins where the bulk of their satellite counts factor in.\ These trends are seen regardless of star-forming definition and are likely a sign of the underlying difference in the observed luminosity/mass functions of these two satellite samples as described by \citetalias[][]{ELVESI}.\ 

\begin{figure*}
	\includegraphics[width=\textwidth]{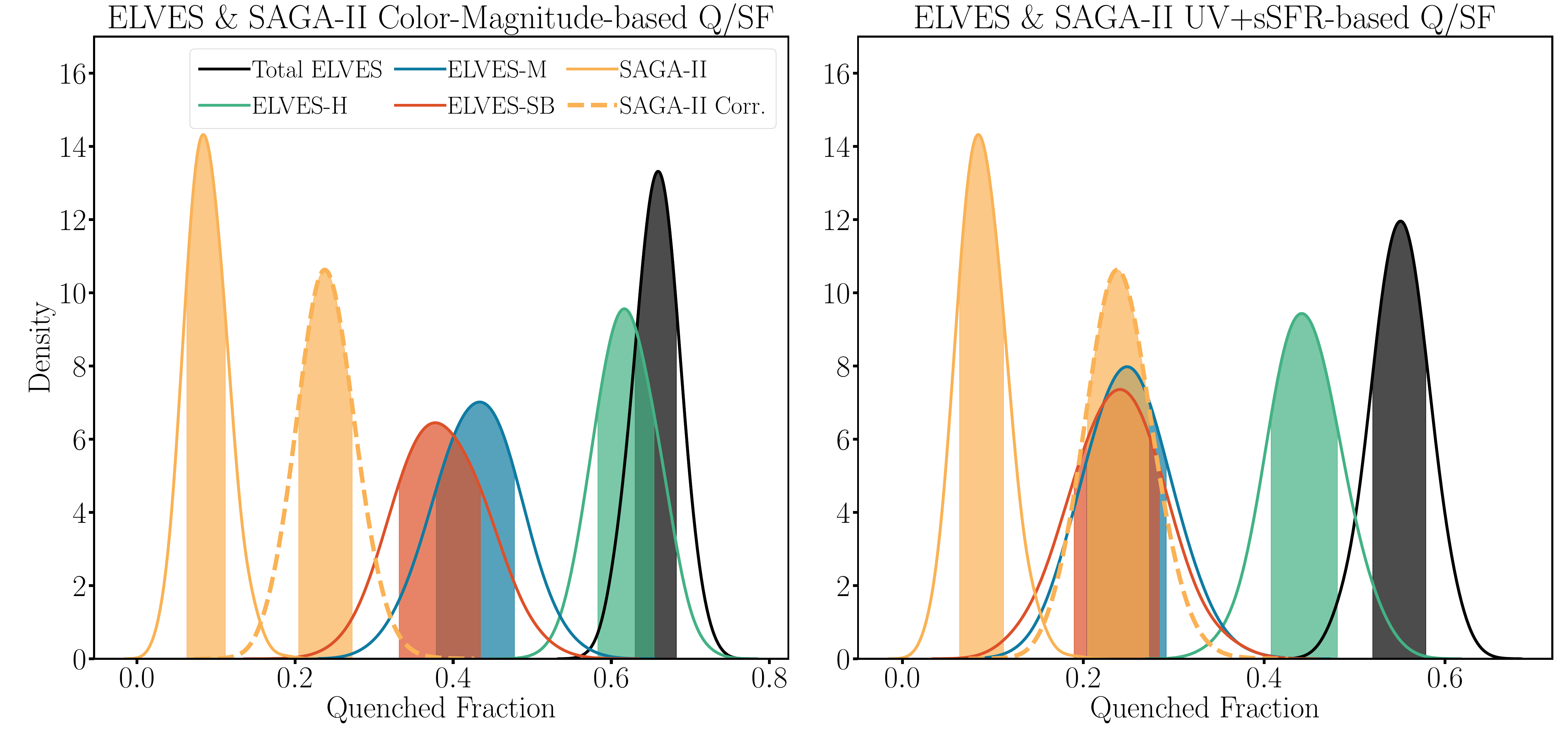}
    \caption{Kernel density estimates of quenched fraction distributions of satellites from the various ELVES sample subsets in comparison to the SAGA--II sample.\ The subsets are coloured above.\ We include two new samples for reference here: the dashed-orange distribution shows the SAGA--II sample ``corrected'' for incompleteness (see text) and the black distribution shows the Total ELVES sample with no cuts applied.}
    \label{fig:qfdensity}
\end{figure*}

The broad agreement between the cumulative number of satellites between the ELVES-M, ELVES-SB, and SAGA--II samples is further solidified when considering various incompleteness/interloper corrections.\  \citetalias{2021Mao} suggest that between $6.6<\mathrm{log}(M_*)<7.8$, $\sim0.7$ satellites are missing per host ($\sim$24 overall) in their characterization of the incompleteness/interloper corrections for the SAGA--II survey.\ Applying these corrections can readily explain the differences in the cumulative number of satellites between the two surveys.\ Indeed, we demonstrate this in Figure \ref{fig:qfdensity} where we show the quenched fraction distributions for the ELVES and SAGA--II samples.\ We derive these distributions by considering the samples as a whole.\ We perform 1000 random draws from each sample with their size equal to that listed in Table \ref{tab:selection}, computing the quenched fraction for each draw.\ With this we are left with a quenched fraction distribution for each sample considered.\ We show the kernel density estimates of these distributions and show their standard deviations $(1\sigma)$ as the shaded regions.\ We include two additional samples for comparison in this analysis.\ First, we show the total ELVES sample without any selection cuts applied in black.\ Second, we show the SAGA--II sample corrected for incompleteness as the dashed orange distribution.\ We incorporate the SAGA--II incompleteness corrections by including the additional 24 satellites in this analysis and assume they are all quenched.\ We can see that the ELVES-M, ELVES-SB, and SAGA--II corrected distributions derived using the sSFR definition agree within their $1\sigma$ regions, while those from the colour-magnitude do not.\ Crucially, in both panels we can also see that there is no significant difference in the quenched fraction distributions between the ELVES-M and ELVES-SB samples.\ This suggests that the surface brightness cut results in insignificant changes to the overall quenched fraction and discuss this further below.\

It should also be noted that there is a strong correspondence in the ELVES sample between the colour-magnitude relation and a morphology-based classification since the former is derived using the latter.\ While we do not focus our comparisons using this morphological classification or a simple NUV detection classification as in \citet{2021Karunakaran}, we show our main results using these methods in Appendix \ref{sec:append}.\

\begin{figure*}
	\includegraphics[width=\textwidth]{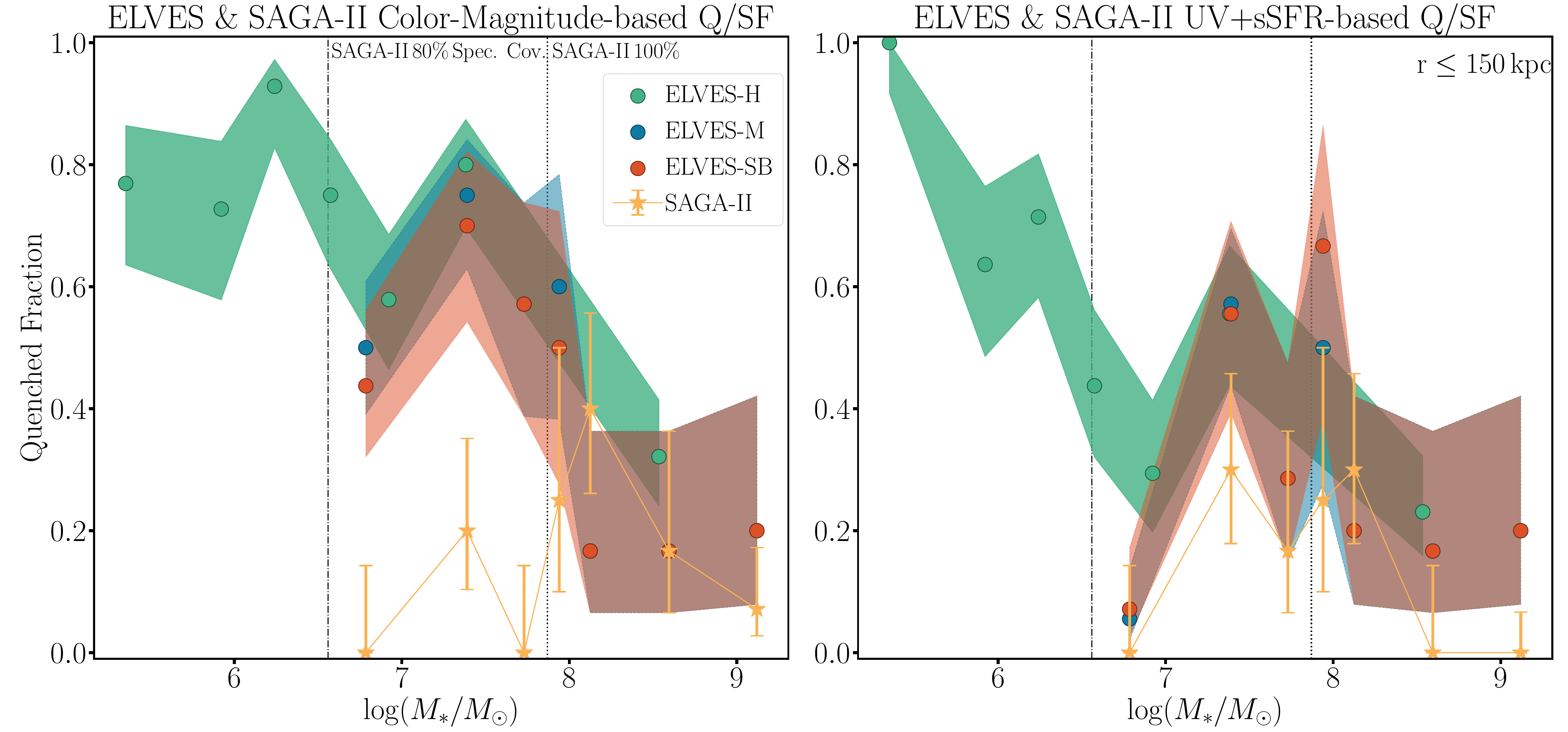}
    \caption{Same as Figure \ref{fig:qf} but only including satellites within 150 kpc of their hosts.
    }
    \label{fig:qf150}
\end{figure*}

\begin{figure*}
	\includegraphics[width=\textwidth]{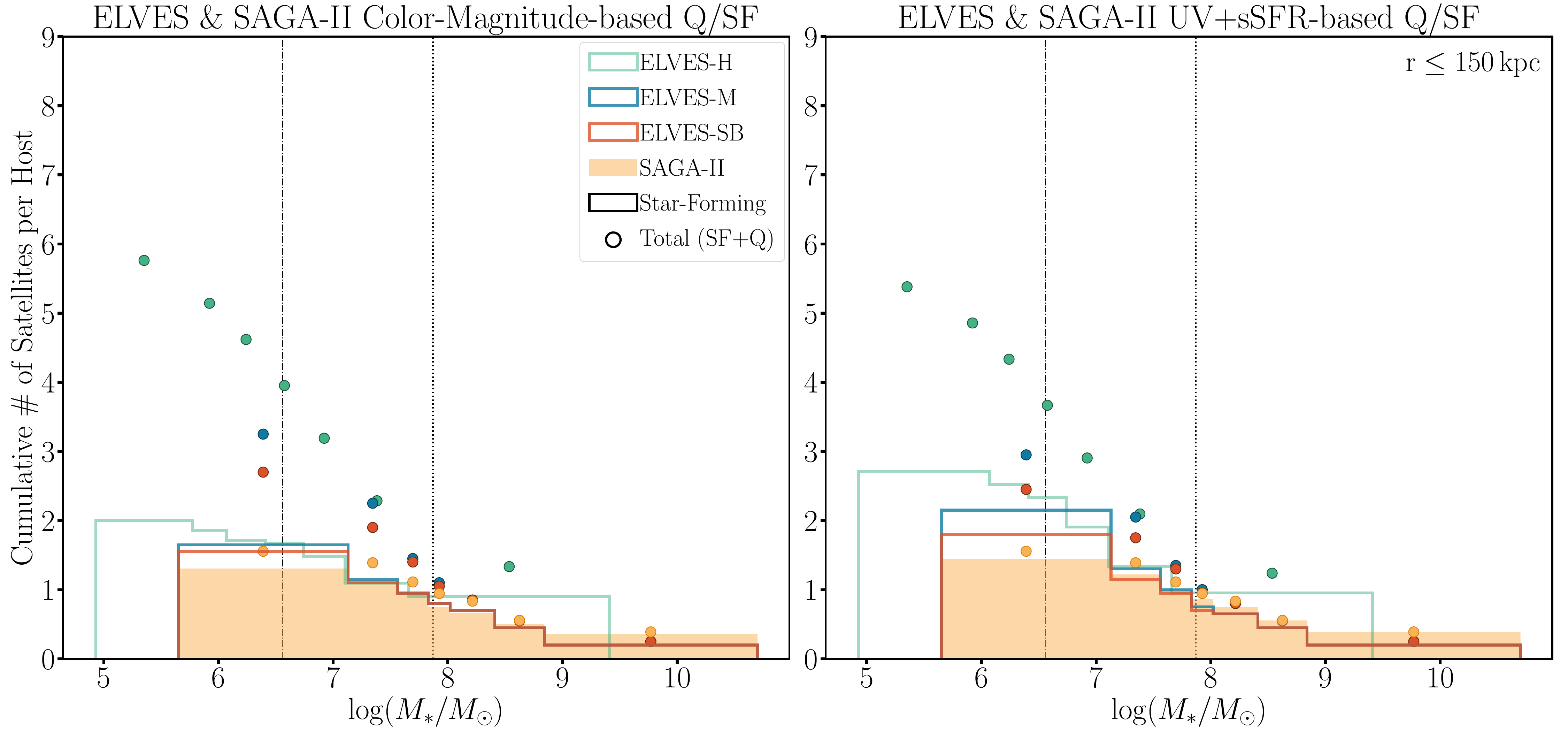}
    \caption{Same as Figure \ref{fig:chist} but only including satellites within 150 kpc of their hosts.}
    \label{fig:chist150}
\end{figure*}

A factor that may play a role in the underlying differences between these two samples is the non-uniform spatial coverage of the ELVES sample.\ As mentioned in Section \ref{sec:sample}, all ELVES hosts are surveyed out to 150 kpc, while most are surveyed to at least 300 kpc.\ We follow a similar methodology to \citetalias[][]{ELVESI} to equate these samples and only select satellites that projected within 150 kpc of their hosts.\ In Figures~\ref{fig:qf150}~and~\ref{fig:chist150}, we show the quenched fractions and the cumulative number of satellites per host as functions of stellar mass, respectively.\ We note that the SAGA--II quenched fractions in Figure \ref{fig:qf150} do not include incompleteness/interloper corrections since they were derived for the full survey coverage.\ Nevertheless, across both star-forming definitions, we see a marginal increase in the quenched fractions, on average (see Table \ref{tab:selection}), as well as an increase in scatter due to fewer satellites within each bin.\ This general increase may have been expected based on the underlying environmental influence on the quenching process.\ We can see that the total number of SAGA--II satellites within 150 kpc is lower than the ELVES sub-samples relative to Figure \ref{fig:chist}.\ This difference in satellite spatial density has been highlighted by \citetalias[][]{ELVESI}.\ However, if we focus on just the star-forming satellites (open and filled histograms), we see that at intermediate stellar masses there is a nearly identical number of star-forming satellites per host.\ The agreement in this region should be considered significant since both samples are sufficiently complete in this mass region.\

Applying a surface brightness criterion, in addition to the SAGA--II absolute magnitude cut ($M_r$$<$$-$12.3 mag), to the ELVES sample leads to marginal differences in the quenched fraction (Figure \ref{fig:qf}) and slightly larger differences in the cumulative number of satellites per host (Figure \ref{fig:chist}) compared to the absolute magnitude criterion alone.\ Based on the more significant decreases seen in the ARTEMIS quenched fractions $(\sim10\%-40\%)$, it is, at first glance, curious that similar decreases are not seen in the ELVES sample.\ Indeed, using satellites from TNG50, \citet{2023Engler} demonstrate that the addition of a surface brightness cut in addition to a magnitude cut does not result in a significant effect to the quenched fractions, particularly at intermediate satellite masses.\ Observationally, Figure \ref{fig:photpropcomp} shows that the vast majority of the LSB satellites are dropped by applying the absolute magnitude cut alone.\ One implication of the minimal effect from the application of the surface brightness criterion is that the SAGA--II photometric catalogs may not be missing as many satellites as initially suggested by \citet{2021Font}.\ Instead, their photometric catalogs are relatively complete and given their faint and/or diffuse nature, these satellites may be missed during their spectroscopic follow-up, as explored by \citetalias[][]{2021Mao}.\

\subsection{A Brief Exploration of Radial Trends}

\begin{figure*}
	\includegraphics[width=\textwidth]{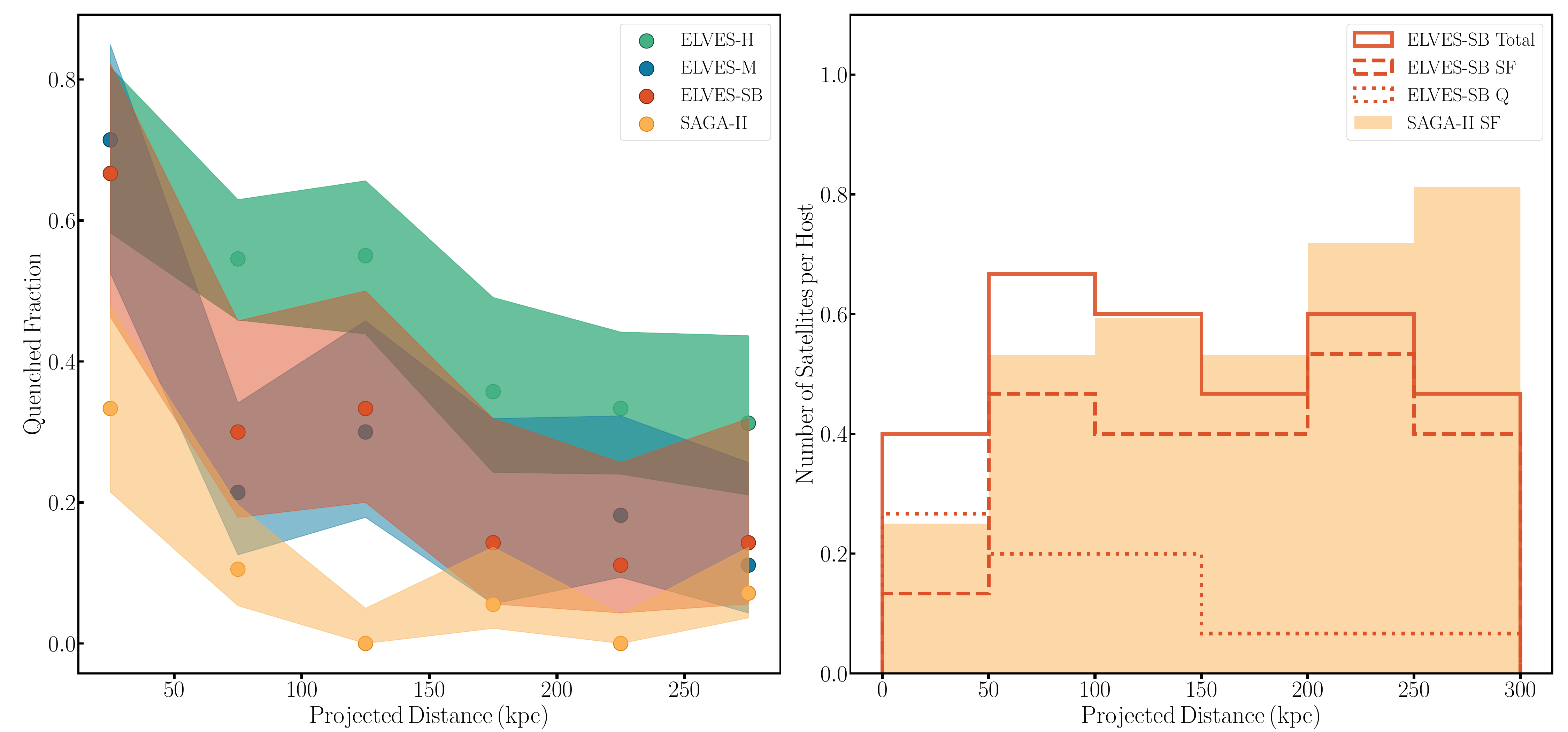}
    \caption{Radial trends for the ELVES and SAGA--II samples.\ ELVES hosts in this comparison are selected to have coverage out to 300 kpc.\ Left: Quenched fractions, computed using the UV+sSFR definition, as a function of projected distance.\ The colour scheme is the same as in the preceding figures (see also Table \ref{tab:selection}).\ A clear environmental effect on the quenched fraction is present in all three ELVES subsets and even marginally in the SAGA--II sample.\ Right: Number of satellites per host as a function of projected distance from their host for the SAGA--II star-forming satellites and the ELVES-SB subset.\ The latter is separated into total (red solid-line), star-forming (red dashed-line), and quenched (red dotted-line) histograms.\ }
    \label{fig:qf_rad}
\end{figure*}

In Figure \ref{fig:qf_rad}, we show the quenched fraction (left) and the number of satellites per host (right) as a function of radius for the ELVES and SAGA--II samples.\ In this comparison, we have restricted the ELVES hosts to those with survey coverage out to 300 kpc.\ We adopt the UV+sSFR star-forming definition given its relatively better agreement between the ELVES and SAGA--II samples.\ From the left panel, we can clearly see the environmental dependence on the quenched fraction as it decreases going from low to high projected distances in all three ELVES sub-samples and is also present for the SAGA--II satellites.\ We can also see that, as before, there is a marginal difference in the quenched fraction radial profiles between the ELVES-M and ELVES-SB samples.\ Although we do not include the SAGA--II incompleteness/interloper corrections, the ELVES samples have, on average, higher quenched fractions across each radial bin.\ However, it is clear to see that SAGA--II, ELVES-M, and ELVES-SB samples are within reasonable agreement at larger project radii ($>150$ kpc) despite not including the aforementioned corrections to the SAGA--II sample.\

It is promising that the average values for the quenched fractions in the ELVES sub-samples $(\sim45\%-60\%)$ are broadly consistent with the quenched fractions within similar radii from the Auriga \citep{Simpson2018} and Latte/FIRE-2 \citep{2022Samuel} simulation suites.\ Interestingly, these quenched fractions appear to reach relatively low values near the virial radius (i.e.\ in the final radial bin) compared to those from these aforementioned simulations.\ However, we believe that this is a result of the different sampling and binning methods and not necessarily a bona fide difference in the radial profiles between the observations and simulations.\

We explore these differences in radial trends from a different point of view in the right panel of Figure \ref{fig:qf_rad} by looking at the satellite number per host as a function of radius.\ We focus this comparison on the ELVES-SB sample, separating the total satellite count (solid red line) into star-forming (dashed red line histogram) and quenched (dotted red line) constituents.\ Also, we only include the SAGA--II star-forming satellites (filled orange histogram).\ We perform a two-sample Kolmogorov-Smirnov (KS) test if the star-forming ELVES-SB and SAGA--II distributions in Figure \ref{fig:qf_rad} (right) are identical and adopt a confidence level of 95\%.\ The KS test results in a \textit{p-}value=0.14.\ Therefore, we cannot reject the null hypothesis that these two samples are drawn from the same distribution.\ Qualitatively, we see that the number of quenched satellites per host in the ELVES-SB sample decreases as a function of projected distance, while the number of star-forming satellites generally increases.\ The star-forming satellites in the ELVES-SB sample hover around $\sim0.4$ satellites per host out to 300 kpc, while the SAGA--II ones continue to increase up to $\sim0.8$.\ This could be a result of the various selection criteria that have been applied to the ELVES sample.\ Indeed, when conducting this comparison with the colour-magnitude relation (or morphology) as the star-forming definition, we see a minor increase in the number of star-forming satellites per host in the larger radial bins and a similar increase for the quenched ones at lower radii.\ However, these minor bin-to-bin differences are difficult to reconcile as physical differences due to the limited sample size considered here, particularly in the more restricted ELVES sub-sample.\

This consistency in the number of star-forming satellites suggests that the apparent dearth of quenched satellites in the SAGA--II survey may not solely stem from bias in their photometric catalogs.\ That is to say, the underlying photometric catalogs may contain satellite candidates that are difficult to spectroscopically confirm due to their faint and/or diffuse nature, as suggested by \citetalias{2021Mao}.\ It is likely that these types of biases may be better constrained in the complete SAGA survey with increased host and satellite sample sizes.\

\section{Summary}\label{sec:summary}
We have presented a comparison of the quenched and star-forming satellite population in the ELVES and SAGA--II samples.\ We applied cuts in host luminosity, satellite luminosity, and satellite surface brightness to the ELVES sample to understand their effects in our comparisons with the SAGA--II satellites.\ With these different sub-samples of ELVES satellites, we compared the quenched fraction and number of star-forming satellites between these two surveys while also testing how different star-forming definitions can affect these comparisons.\ Our primary conclusions from this work are as follows:

\begin{itemize}
    \item We find that the quenched fractions, calculated in a consistent manner using a UV-derived sSFR, in the ELVES and SAGA--II samples are in agreement after applying either (i) the SAGA--II absolute magnitude cut or (ii) a surface brightness cut in addition to the SAGA--II absolute magnitude cut to the ELVES sample (see Figure \ref{fig:qf}).\ Additionally, there is an even stronger agreement between ELVES and SAGA--II when only accounting for satellites within 150 kpc (see Figure \ref{fig:qf}).\ This lends confidence to these quenched fraction estimates as the underlying samples of satellites are found via different techniques.
    \item The absence of any significant effect on the ELVES sample after applying an additional surface brightness cut implies that the SAGA--II sample is not missing additional satellites beyond what they have already accounted for in their incompleteness corrections.\
    \item Similarly, we find broad agreement in the cumulative number of all satellites between the ELVES and SAGA--II samples, regardless of star-forming definition, particularly when the SAGA--II incompleteness corrections are accounted for.\
    \item Our brief investigation of radial trends in both the ELVES and SAGA--II samples found that (i) as expected, the quenched fraction decreases as a function of radius and (ii) there is broad agreement in the number of star-forming satellites in the ELVES and SAGA--II samples as a function of radius.\
    \item At first glance, there may be tension between the quenched fraction seen in simulations, which use the Local Group as a benchmark, and larger statistical samples of Milky Way-like systems.\ Additionally, the Local Group itself may be an outlier among Milky Way-like systems.\ Continuing to expand observational samples with multiwavelength datasets is crucial to better elucidate these tensions.\ Similarly, \citet{2023Engler} demonstrate the utility of applying numerous observational selection criteria to large simulated samples.\ Having additional simulation samples apply similar rigor can readily aid in understanding these potential tensions, such as the quenched fractions of satellites. 
\end{itemize}

The work we have presented here expands upon several investigations of the quenched fraction around Milky Way-like systems.\ However, more stringent constraints on these properties should be made using spectroscopic follow-up to better characterize the association of these satellites (i.e.\ are they gravitationally bound to their hosts? Are they potential backsplash systems?).\ Additionally, comprehensive studies of their star-forming nature via resolved H$\alpha$ and UV imaging (e.g.\ 20\% of ELVES satellites do not have UV imaging) would be valuable extensions that could further elucidate the results presented here.\ HI observations are another crucial element that could shed light on both quenching mechanisms (i.e. ram pressure stripping) and the star-forming potential of these faint satellite systems; some of this work is already underway \citep{2022Karunakaran}.\ 

\section*{Acknowledgements}
We thank the referee for their useful comments and suggestions to improve the quality of this paper.\ AK acknowledges financial support from the State Agency for Research of the Spanish Ministry of Science, Innovation and Universities through the "Center of Excellence Severo Ochoa" awarded to the Instituto de Astrof\'{i}sica de Andaluc\'{i}a (SEV-2017-0709) and through the grant POSTDOC$\_$21$\_$00845 financed from the budgetary program 54a Scientific Research and Innovation of the Economic Transformation, Industry, Knowledge and Universities Council of the Regional Government of Andalusia.\ KS acknowledges support from the Natural Sciences and Engineering Research Council of Canada (NSERC).\ BMP is supported by an NSF Astronomy and Astrophysics Postdoctoral Fellowship under award AST2001663.\ Research by DC is supported by NSF grant AST-1814208.\ DJS acknowledges support from NSF grants AST-1821967 and 1813708.

This work made use of \textsc{astropy} \citep{2013Astropy,2018Astropy}, \textsc{jupyter} \citep{2021Jupyter}, \textsc{matplotlib} \citep{2007matplotlib}, \textsc{numpy} \citep{2020Numpy}, \textsc{pandas} \citep{2010pandas}, \textsc{seaborn} \citep{2021seaborn}, \textsc{scipy} \citep{2020SciPy}, and \textsc{statsmodels} \citep{2010statsmodels}.

This research made use of data from the SAGA Survey (sagasurvey.org). The SAGA Survey was supported by NSF collaborative grants AST-1517148 and AST-1517422 and by Heising–Simons Foundation grant 2019-1402.
\section*{Data Availability}
The data used in this work are publicly available from the original publications describing these samples.\

\bibliographystyle{mnras}
\bibliography{references} 
\newpage
\appendix
\section{Classifying Satellites Using Morphology and UV-detection}\label{sec:append}
Here we show the quenched fractions and number of satellites for the ELVES sample derived using their morphological classifications.\ For the SAGA--II satellites, we adopt the simple star-forming criterion, i.e.\ $(S/N)_{NUV}>2$, from \citet{2021Karunakaran}, where they found a high correspondence of satellites with both H$\alpha$ and NUV emission.\ We use this simple NUV detection definition only in comparison to the ELVES morphological classification, analogous to the comparison between morphology- and H$\alpha$-based quenched fractions in \citetalias{ELVESI} (see their Figure 11).\ 

\begin{figure*}
	\includegraphics[width=\columnwidth]{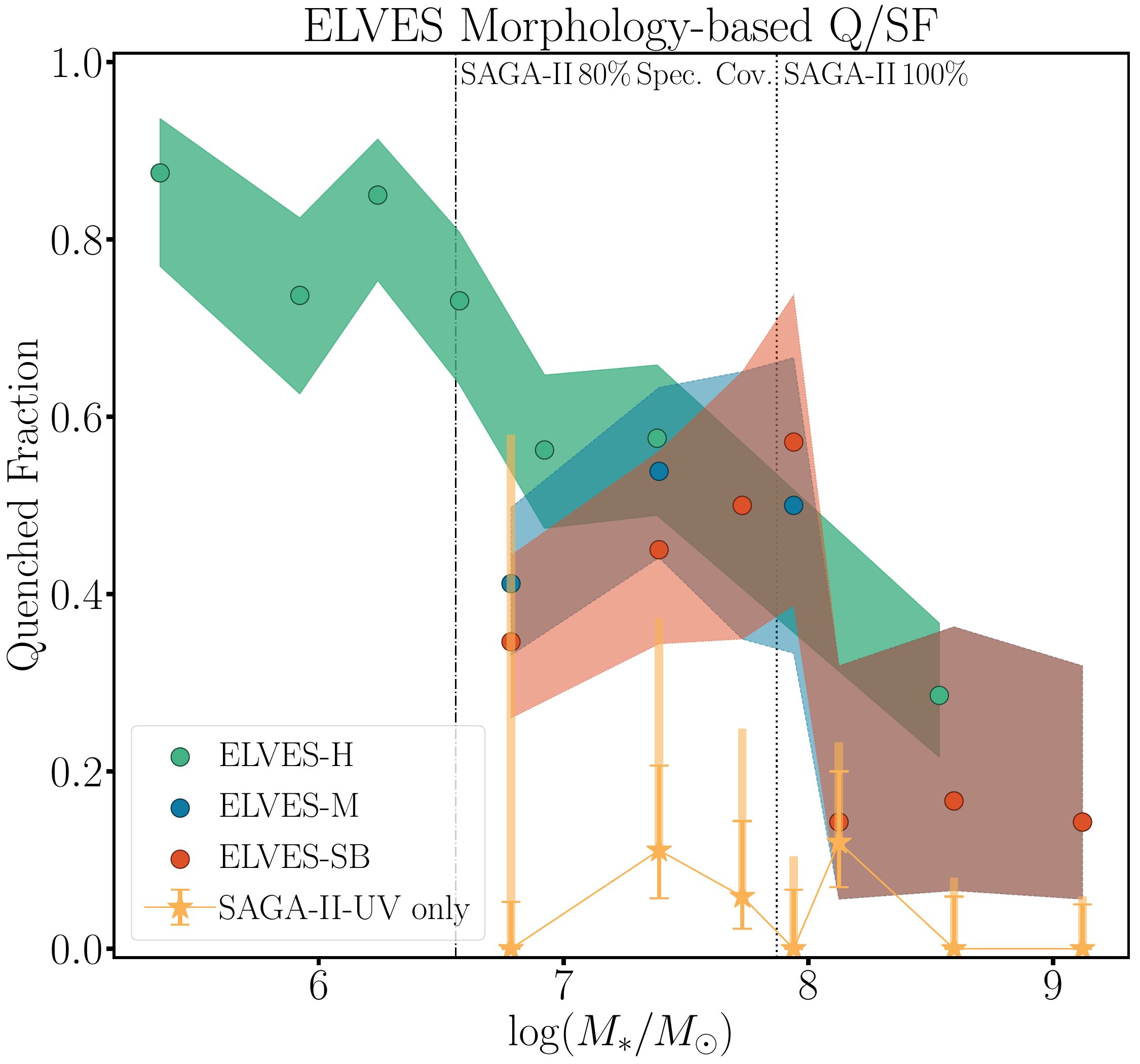}
        \includegraphics[width=\columnwidth]{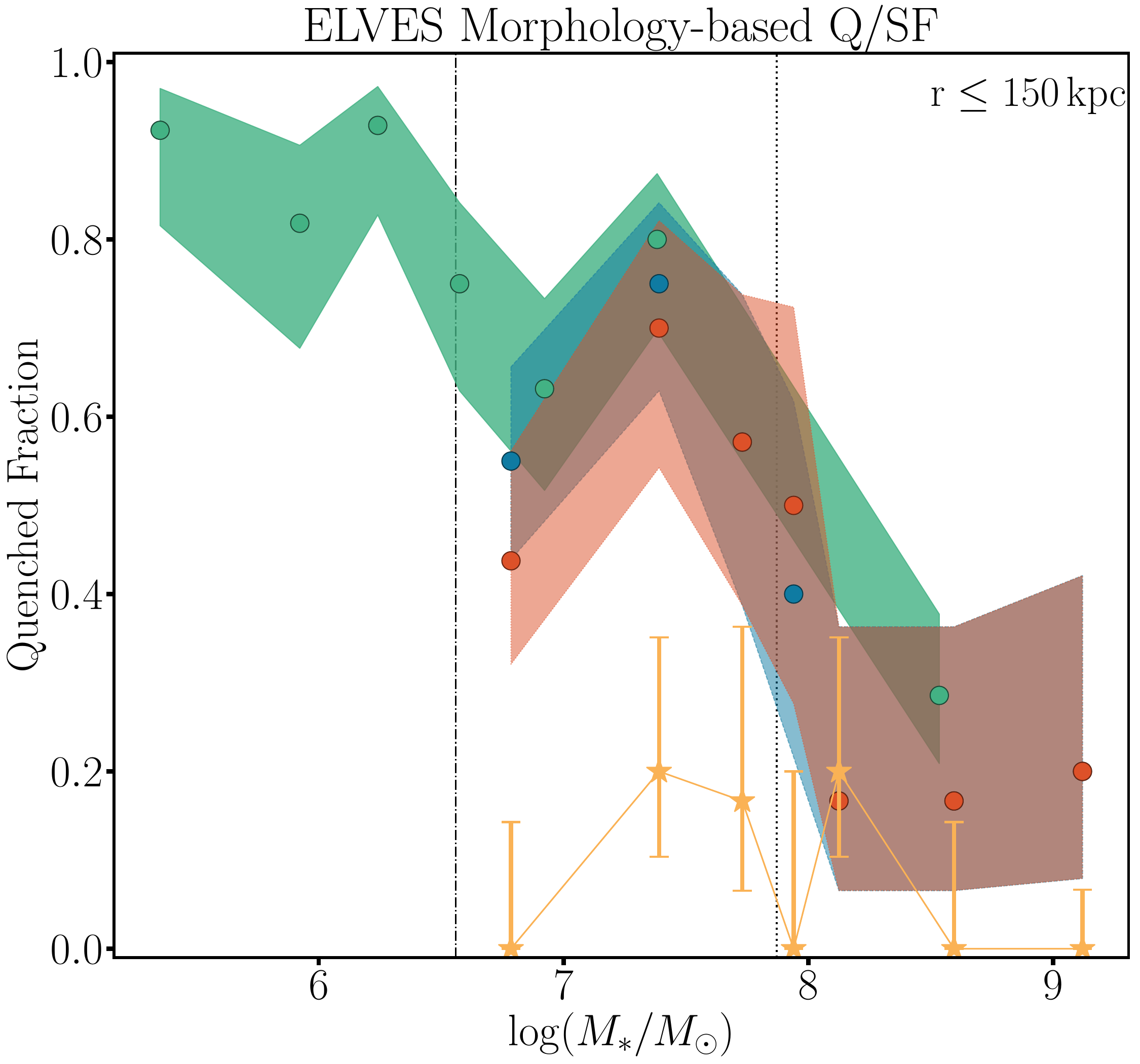}
    \caption{Same as Figures \ref{fig:qf} and \ref{fig:qf150} for all satellites (left) and just those within 150 kpc (right) but using morphology as the star-forming classification for the ELVES sample and significant NUV emission as the star-forming classification for SAGA--II.
    }
    \label{fig:qfmorphuv}
\end{figure*}

\begin{figure*}
	\includegraphics[width=\columnwidth]{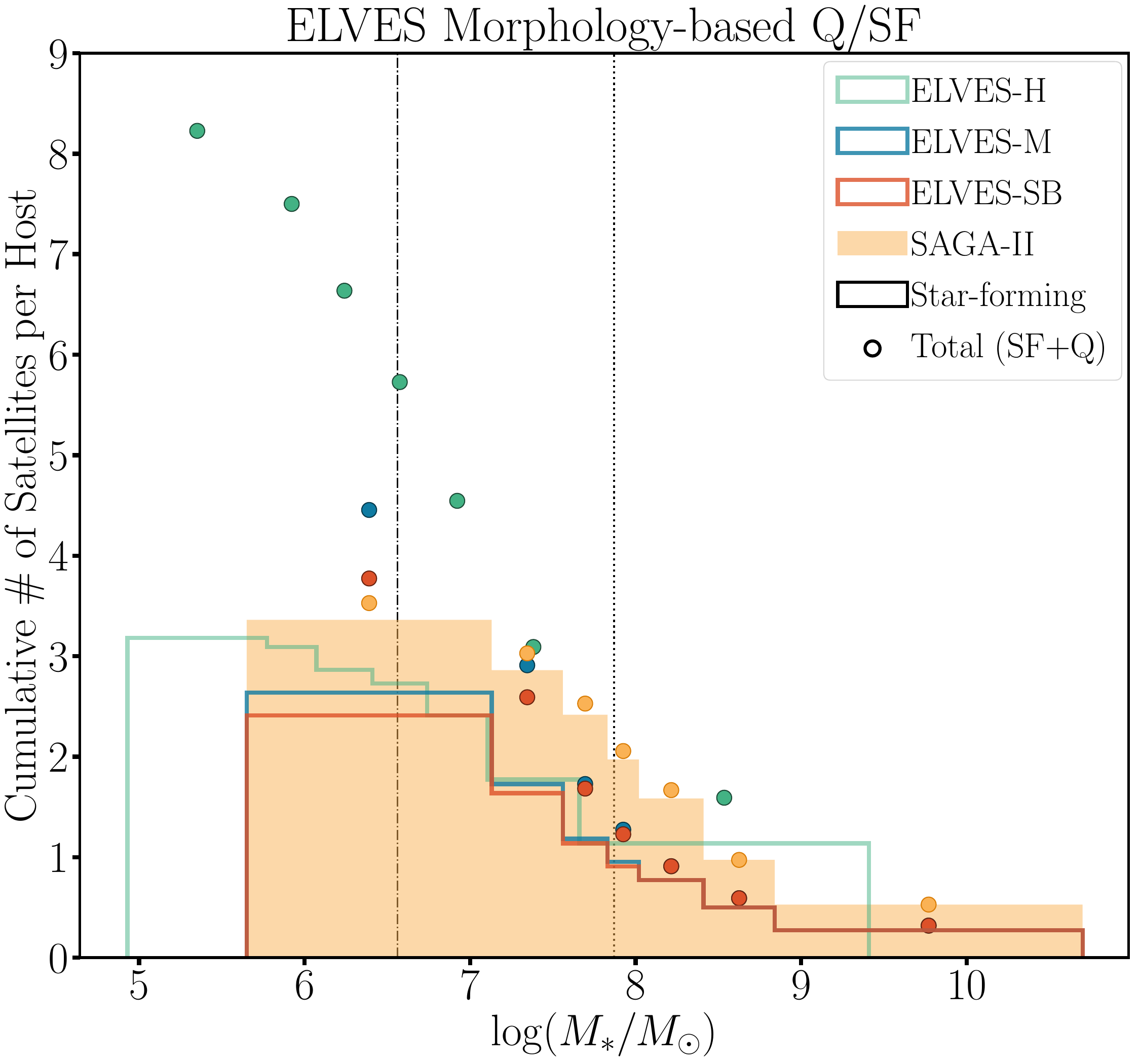}
 	\includegraphics[width=\columnwidth]{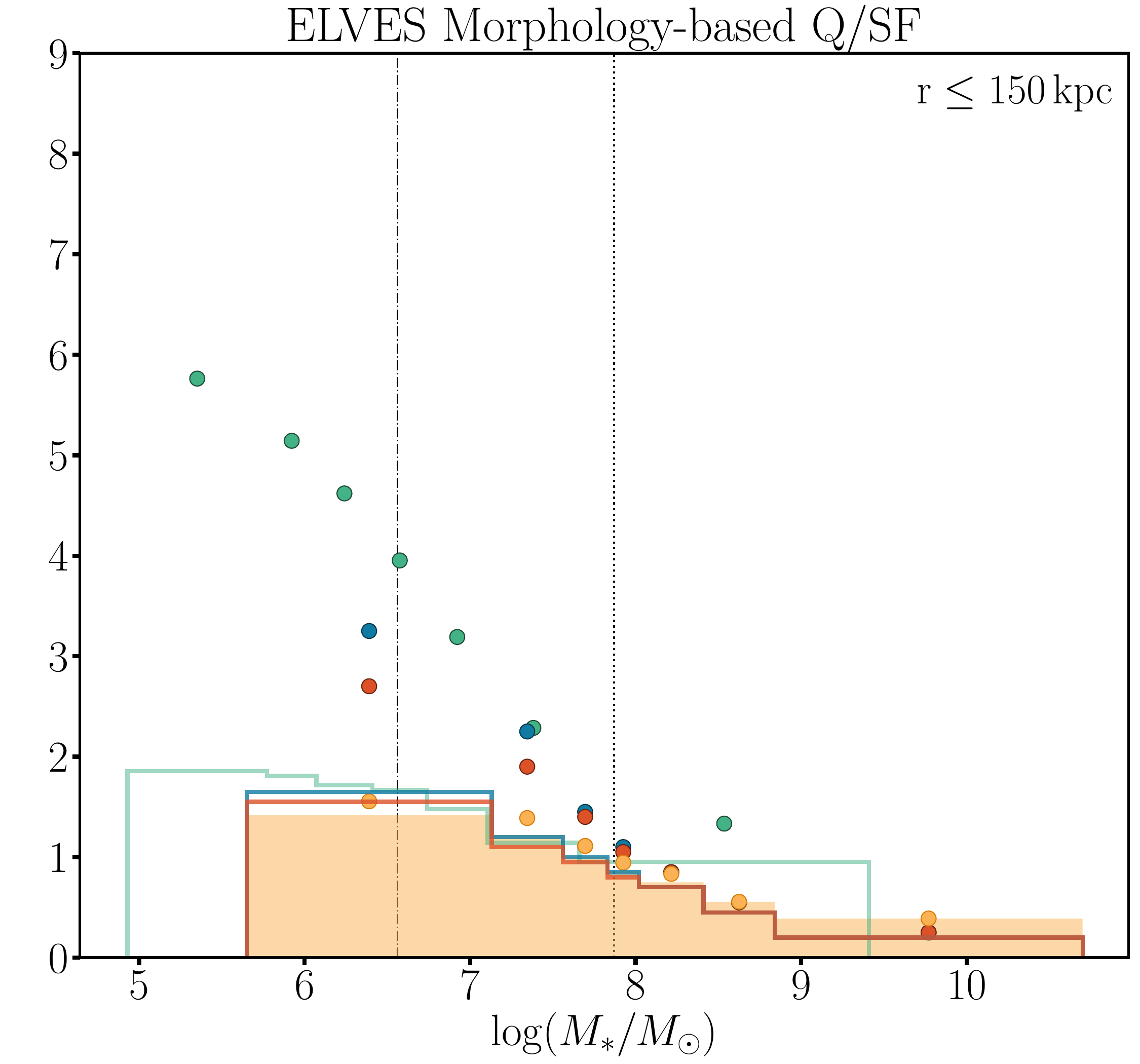}
    \caption{Same as Figure \ref{fig:chist} and \ref{fig:chist150} for all satellites (left) and just those within 150 kpc (right) but using morphology as the star-forming classification for the ELVES sample and significant NUV emission as the star-forming classification for SAGA--II.}
    \label{fig:chistmorphuv}
\end{figure*}

Comparing the ELVES quenched fractions and the number of satellites from Figures \ref{fig:qfmorphuv} and \ref{fig:chistmorphuv}, we can see there is a negligible difference between the classifications based on the morphology and the colour-magnitude relation regardless of the selection criteria applied (ELVES-H, -M, or -SB).\ This consistency between these two classification methods should be expected given the premise of the colour-magnitude relation determination in \citet{Carlsten2021}.\ It is interesting to point out in there is an even tighter correspondence between the morphology-based and colour-magnitude relation-based classifications when considering satellites within 150 kpc.\

% Don't change these lines
\bsp	% typesetting comment
\label{lastpage}
\end{document}